\documentclass[12pt,preprint]{aastex}

\newcommand{\lsim}{\raisebox{-0.3ex}{\mbox{$\stackrel{<}{_\sim} \,$}}}

\def\gta{\ifmmode {\mathbin{\lower 3pt\hbox   
    {$\,\rlap{\raise 5pt\hbox{$\char'076$}}\mathchar"7218\,$}}}
    \else {${\mathbin{\lower 3pt\hbox
    {$\rlap{\raise 5pt\hbox{$\char'076$}}\mathchar"7218\,$}}}
    $}\fi}
\def\lta{\ifmmode {\,\mathbin{\lower 3pt\hbox   
    {$\,\rlap{\raise 5pt\hbox{$\char'074$}}\mathchar"7218\,$}}}
    \else {${\mathbin{\lower 3pt\hbox
    {$\rlap{\raise 5pt\hbox{$\char'074$}}\mathchar"7218\,$}}}
    $}\fi}

\shorttitle {Spectral and timing behavior of 1A 1744--361}
\shortauthors {Bhattacharyya et al.}

\begin{document}

\title {RXTE Observations of 1A 1744--361: Correlated Spectral and Timing Behavior}

\author {Sudip Bhattacharyya\altaffilmark{1,2}, Tod
E. Strohmayer\altaffilmark{2}, Jean H. Swank\altaffilmark{2}, and
Craig B. Markwardt\altaffilmark{1,2}}

\altaffiltext{1}{Department of Astronomy, University of Maryland at
College Park, College Park, MD 20742-2421}

\altaffiltext{2}{X-ray Astrophysics Lab,
Exploration of the Universe Division,
NASA's Goddard Space Flight Center,
Greenbelt, MD 20771; sudip@milkyway.gsfc.nasa.gov,
stroh@clarence.gsfc.nasa.gov, swank@milkyway.gsfc.nasa.gov, craigm@milkyway.gsfc.nasa.gov}

\begin{abstract}

We analyze Rossi X--ray Timing Explorer ({\it RXTE}) Proportional Counter
Array (PCA) data of the transient low mass X--ray binary (LMXB) system 1A 1744--361.
We explore the X--ray intensity and spectral evolution of the source, perform
timing analysis, and find that 1A 1744--361 shows `atoll'
behavior during the outbursts. The color-color diagram
indicates that this LMXB was observed in a low intensity spectrally hard (low-hard) state
and in a high intensity `banana' state. The low-hard state shows a horizontal
pattern in the color-color diagram, and the previously reported `dipper QPO'
appears only during this state. We also perform energy spectral analyses, and
report the first detection of broad iron emission line and iron absorption edge 
from 1A 1744--361.

\end{abstract}


\section {Introduction} \label{sec: 1}

The transient low mass X--ray binary (LMXB) 1A 1744--361 was discovered by Ariel V in 1976, when
it was in outburst (Davison et al. 1976; Carpenter et
al. 1977). Since then four more outbursts have been observed from this
source in the years 1989, 2003, 2004 and 2005. Bhattacharyya et al. (2006) found an
unambiguous thermonuclear X--ray burst from the 2005 {\it RXTE} PCA data of
this source, which confirmed the suggestion of Emelyanov et al. (2001) that 
this source harbors a neutron star. This burst also showed millisecond period brightness 
oscillations during the intensity rise (with the frequency $\sim 530$ Hz; Bhattacharyya 
et al. 2006), which gave the spin frequency of the neutron star ($\sim 530$ Hz).
This is because, these oscillations are produced by an asymmetric brightness pattern on the
stellar surface that is modulated by rotation of the star (Strohmayer et al. 1996; Chakrabarty
et al. 2003; Strohmayer \& Bildsten 2006). Bhattacharyya et al. (2006) also
found energy dependent dips in the 2003 PCA data, which established that this source
is a dipping LMXB (dipper). Such LMXBs exhibit modulation of soft X-ray intensity
with the binary orbital period (which gave the orbital period $\sim 97$ min of 1A 1744--361;
Bhattacharyya et al. 2006). This modulation is believed to be caused
by structures above the accretion disk (White \& Swank 1982). This is
possible only if the dippers are high inclination systems, so that the
line of sight passes through these structures. Therefore, dippers provide
an opportunity to constrain the properties of the upper layers of accretion disks (and the
photoionized plasma above them; Jimenez-Garate et al. 2003) in LMXBs.
Moreover, recently spectral lines have been discovered from several dippers, 
such as EXO 0748-67 (Cottam et al. 2001), 
XB 1916--053 (Boirin et al. 2004), X 1624--490 (Parmar et al.  2002), etc.
This shows that 1A 1744--361 might be a promising source to search for these 
features, which could be 
useful to understand various X--ray emitting and absorbing components of LMXBs.

It is also important to determine the broad spectral and timing category 
(e.g., Z, atoll, weak LMXB, etc.; see, for example, 
van der Klis 2004; Kuulkers et al. 1997; Wijnands et al. 1998; Bhattacharyya 2006) 
of 1A 1744--361 in order to understand the nature of this source.
Z sources are the most luminous LMXBs and trace `Z'-like curves 
(with three branches: horizontal, normal and flaring) in the 
color-color diagram. The ordinary atoll sources have luminosities in the range
$\sim 0.01-0.2 L_{\rm Edd}$ ($L_{\rm Edd}$ is the Eddington luminosity), and trace 
`C'-like curves (with two branches: low intensity `island' and high intensity `banana';
van der Klis 2004).
The various portions of these curves (combined with the correlated timing features)
indicate different source states (van der Klis 2004). Recently, a low intensity
spectrally hard (low-hard) state has been observed from several atoll sources.
Because of its horizontal branch like pattern in the color-color diagram,
some authors (Muno, Remillard, \& Chakrabarty 2002 (hereafter MRC);
Gierli\'nski \& Done 2002) suggested that this state might be to some extent similar
to the horizontal branch state of Z sources. Such a picture, if true, favors at least partial
unification of Z and atoll sources. However, this horizontal branch of atoll sources
occurs at a much lower luminosity, much harder spectral state, and with a much
longer time scale (compared to those of Z sources; MRC), and hence we
call this state `atoll horizontal branch' (AHB). Here we note that AHB might 
actually be the known extreme island state (EIS; van der Klis 2004) of atoll sources, and detailed
study of the low intensity states of the atoll sources is necessary to resolve 
this. We also note that there is another category of LMXBs (known as weak LMXB;
luminosity $ \lsim 0.01 L_{\rm Edd}$),
which comprises the faint burst sources, the low-luminosity transients (e.g., 
Heise et al. 1999), etc. Many of
these sources appear to be low-luminosity atoll sources (van der Klis 2004), and
show timing properties similar to those of atoll sources.

In \S~2, we calculate the color-color diagram, perform timing analysis, and show that
1A 1744--361 exhibits atoll behavior during outbursts. 
We also suggest that the previously discovered 
low frequency dipper QPO (Bhattacharyya et al. 2006; Jonker, van der Klis \& Wijnands 1999) is connected
to the low-hard state of the source. In \S~3, we describe the first detection
of broad iron emission line and iron absorption edge from the energy spectra of 1A 1744--361,
and in \S~4 we discuss our results. 

\section {Color-Color Diagram and Timing Analysis} \label{sec: 2}

The transient source 1A 1744--361 was observed with {\it RXTE} PCA during three outbursts.
The duration (year, RXTE proposal number, RXTE observation duration)
of these outbursts were $\sim 2$ months (2003, P80431, $\sim 39$~ks),
$\sim 20$ days (2004, P90058, $\sim 2$~ks), and $\sim 40$ days (2005, P91050, $\sim 15$~ks).
In Fig. 1, we show the light curves from the ASM and from the PCA scans of the 
galactic bulge (Swank \& Markwardt 2001), together with the time intervals of the pointed 
observations with {\it RXTE}.
Here each interval is during an outburst, and contains several observations.
We note that the scans across the galactic bulge are modeled in terms of source
contributions and an ellipsoidal estimate of the unresolved flux in the galactic ridge.
This analysis gives a best fit to the flux of 1A 1744--361 when it is
in quiescence, as well as when it is in outburst. Although, this procedure produces an
estimate of the quiescent flux of the source at the level of a few tenths of a mCrab, 
we do not consider it a `positive detection' of 1A 1744--361 in quiescence, because 
of systematic errors. The source was not observed with {\it ASCA}, {\it BeppoSAX},
or {\it XMM-Newton}. However, {\it Chandra} observed the field of the Ariel V error
circle while 1A 1744--361 was in quiescence (Torres et al. 2004), and a faint 
X-ray source in this field was identified with the optical source and the radio
source reported by Steeghs et al. (2004) and Rupen, Dhawan, \& Mioduszewski (2003)
respectively. This showed that this faint X-ray source was 1A 1744--361 in quiescence,
and the very low source flux (13 counts in $\sim 16$ ks; Torres et al. 2004) 
was below the PCA detection level.

Fig. 1 shows that the first (2003) outburst was the strongest and the longest,
and it had two peaks. The 2005 outburst was also strong, but the 2004 outburst was weak
and provides an opportunity to understand the source in a low-hard state (see later).
The pointed observations were made near the first peak of the 2003 outburst, and
near the peaks of the other two outbursts.
We have computed a color-color diagram (CD; Fig. 2) using the {\it RXTE} PCA
(from only the top Xenon layers to increase signal to noise)
pointed observations. These data are from the same gain epoch (epoch 5),
and hence the gains of the Proportional Counter Units (PCUs) are almost the same.
This ensures minimal shifting of the source track in the CD due to the differences of the energy 
boundaries. The PCUs 0, 2 and 3 were on during most of these observations, and we have used the
data from these PCUs to calculate most of 
the CD. However, for a few ObsIDs we have used data from other PCU
combinations. For these ObsIDs, we choose the PCA channel ranges carefully, so that the energy
boundaries are similar to those for other ObsIDs.
We have defined soft and hard colors (used in the CD; Fig. 2) as the ratio of
the background-subtracted counts approximately 
in the $(3.5-5.1)/(2.2-3.5)$ and $(8.5-17.8)/(5.1-8.5)$ keV energy bands respectively.
We have also calculated a hardness-intensity diagram (HID; Fig. 2) using these data with the
background-subtracted intensity in the energy range $\sim 2.2-17.8$ keV.
These definitions are close to the definitions used by 
MRC, and will facilitate the comparison of our results with those of these authors.
We note that we have used the `FTOOLS' command `pcabackest' for estimating
the backgrounds, and have not included contribution due to galactic ridge emission
which is in the $1^{\rm o}$ field of view of the PCA collimator.
This is because the source was always $> 10-100$ times brighter than
the galactic ridge, and hence the ridge contribution (see \S~3) to the colors 
is not significant.

From the CD and the HID, it is clear that 1A 1744--361 was observed in two
distinctly different states: a low intensity spectrally hard (low-hard) state in 2004 and
a high intensity state in 2003 and 2005. 
In order to explore the nature of 1A 1744--361, first we have calculated the typical X--ray fluxes from
this source in these two states by fitting spectra with models. The $2-20$ keV fluxes are
$\sim 4.3 \times 10^{-10}$ and $\sim 3.2 \times 10^{-9}$ ergs cm$^{-2}$ s$^{-1}$ respectively.
The peak $2-20$ keV flux during a non-photospheric-radius-expansion thermonuclear X--ray burst from this
source was $1.9 \times 10^{-8}$ ergs cm$^{-2}$ s$^{-1}$ (Bhattacharyya et al. 2006), which implies that the 
low intensity state luminosity $L_{\rm low}/L_{\rm Edd} \lsim 0.02$ and the high intensity state
luminosity $L_{\rm high}/L_{\rm Edd} \lsim 0.17$. Here, $L_{\rm Edd}$ is the Eddington luminosity.
Both of these luminosities are consistent with those of atoll sources (van der Klis
2004). Here we note that the typical luminosities
of Z sources are close to the Eddington luminosity (van der Klis 2004), and hence much higher
than the luminosity of 1A 1744--361. 

In order to more definitively show that this source is not a Z source,
next we have compared our CD (Fig. 2) with that of MRC.
Fig. 1 of MRC shows that the flaring branches (FBs) of Z sources have hard colors 
close to 0.2 (except for GX 17+2). As the hard colors of 1A 1744--361 in the high intensity state
are close to 0.3, this indicates that if this source is a Z source, it
was not on the FB during the 2003 and 2005 observations. However, during the 2003
observations the source did show flares, which are not usual for the
other two states (normal branch and horizontal branch) of Z sources.
The hard colors of 1A 1744--361 in the low intensity state are close to $5.5-6$. Z sources
normally do not show such high values of hard colors (see Fig. 1 of MRC).
Therefore, it is not likely that 1A 1744--361 is a Z source.

Fig. 1 of MRC shows that the hard colors of atoll sources are always around or greater than 0.3
(except for GX 13+1). This, and especially the observed hard colors of the banana branches of
atoll sources (see MRC) strongly indicate that 1A 1744--361 was in the banana branch during 
2003 and 2005 outbursts (high intensity states; see Fig. 2). Therefore,
1A 1744--361 exhibited atoll behavior during these strong outbursts. Here we note that
during the first observation of 2005 (ObsID 91050-05-01-00), the source was spectrally slightly harder
and less intense ({\it square} symbols in Fig. 2), which might be an indication 
of transition between banana and island states (van der Klis 2004). We also note that this observation
registers the only thermonuclear X--ray burst seen from this source (Bhattacharyya et al 2006; see
also Fig. 2).

The CD (Fig. 2) shows that 1A 1744--361 traced a horizontal pattern at high hard color values in 2004.
This is likely to indicate the AHB state (see \S~1) of the source, because its intensity was very
low, it was spectrally very hard, and it followed a horizontal track in the CD.
Therefore, even during the weak outburst (in 2004), the source properties were consistent 
with those of atoll sources.
Here we note that the small amount of observation does not allow us to determine the duration of this state.
However, the time separation between the two ObsIDs (90058-04-01-00 \& 90058-04-02-00) in
this state was $\sim 40$ hrs., which may indicate that at least during this time
period the source was in this low-hard state. We also note that 1A 1744--361 was not detected 
with hard colors in the range $\sim 0.37-0.5$, and it seems that the future observation of 
these hard colors will be indicative of the island state of the source (van der Klis 2004).

As a source state is characterized by its timing properties in addition to
the position of the source on the CD, we have computed and
fitted the power spectra of 1A 1744--361 (Table 1). We have found that the
high frequency power spectra of all data segments are featureless, and show only
white noise (but see Bhattacharyya et al. 2006 for the description of a possible
kHz QPO). Therefore, for this paper we have computed and fitted the low frequency (up to
$\sim 100$ Hz) power spectra of the data of four representative portions of the
CD (Fig. 3). In these calculations we have used the event mode data from all the available PCUs,
and divided the data into 250 s segments.
We have constructed light curves with a bin size of $1/256$ s (all the event modes
have bin sizes which are submultiples of this), and then performed Fourier transforms
of the time segments.
An average of all these transforms from a data set gave the power spectrum.
This spectrum has been binned geometrically in order to increase the signal to noise.
Fig. 3 shows these spectra (from four data sets), the best fit models, and
the positions of the corresponding data segments on the CD. The best fit parameter values
are given in Table 1. The power spectrum of the low intensity AHB state (2004
data) of 1A 1744--361
can be sufficiently modeled with a constant (describing the white noise) and a
Lorentzian (describing the dipper QPO; Bhattacharyya et al. 2006). The centroid frequency
and the rms of the Lorentzian do not significantly depend on photon energies (within error bars),
which is consistent with the `dipper QPO' interpretation of this feature (Jonker et al. 1999).
The power spectra of all the high intensity states (2003 and 2005 data) of the source
are well fitted with a constant and a power-law. The power-law describes the
`very low frequency noise' (VLFN; van der Klis 2004; Boirin et al. 2000; Berger, \& van der Klis
1998; Agrawal \& Bhattacharyya 2003), which strongly indicates that 1A 1744--361 was
on the banana branch in 2003 and 2005. However, during the first observation of 2005 
(ObsID 91050-05-01-00), the strength of the VLFN was significantly less than that
during the other observations (at high source intensity; see Table 1 and Fig. 3). This
supports the guess (made earlier in this section) that during this 2005 observation, the source was
in a transition state (between banana and island; van der Klis 2004).
From these timing analyses, as well as from the source luminosity and the study of the CD, 
we conclude that 1A 1744--361 exhibits atoll behavior during the outbursts.

\section {Spectral Analysis} \label{sec: 3}

A CD gives an idea about the spectral evolution of the source. But to fully
understand the spectral properties of 1A 1744--361, a detailed spectral analysis
is essential. However, before describing such an analysis, we note that {\it RXTE} PCA
registered excess X--ray emission from the galactic ridge during the pointed
observations of the source. This is because the location of 1A 1744--361
is in the galactic ridge (galactic latitude $\approx -4^{\rm o}.2$), and the
FWHM spatial resolution of the PCA is $1^{\rm o}$ (Valinia \& Marshall 1998). 
Therefore, we need to take this
emission into account during each spectral fitting, especially if we are interested in
iron features in the spectrum (as the unresolved excess emission contains an iron emission line (at
$\sim 6.7$ keV); Koyama et al. 1986). But, the `pcabackest' command (for background
calculation) of `FTOOLS' does not include the galactic ridge contribution (Jahoda et al. 2005).
We should, therefore, model the galactic ridge spectrum at the position of 1A 1744--361
(when the source is not in outburst) and use this model (with the parameters frozen
to the best fit values) in addition to the other model components to describe
the source spectra. 

We had one pointed observation of the source position during the source quiescence.
While this observation was only about 0.9 ks in duration, we believe that the fact that it is 
from exactly the source position makes it preferable to sum of slews over the region.
We have fitted the `pcabackest' background subtracted {\it RXTE} PCA spectrum 
with the absorbed `Raymond-Smith' plus power-law model
({\it wabs*(raymond+powerlaw)}; Valinia \& Marshall 1998). 
The best fit parameters
are given in Table 2. We have used this model (with these best fit parameter values) 
as a model component for all the source spectral analyses.
Next, we have chosen a high intensity source data segment (ObsID 80431-01-02-02) in order
to find out typical spectral properties of 1A 1744--361 during high intensity states. We have fitted
(after `pcabackest' background subtraction) the corresponding spectrum with various
{\it XSPEC} models (see Table 3). The model that gives a reasonable $\chi^2/{\rm dof}$ value,
contains an absorbed Comptonization ({\it compTT} in {\it XSPEC}) plus blackbody
({\it bbodyrad} in {\it XSPEC}) as the continuum, a broad emission line ({\it gauss} in {\it
XSPEC}) and an absorption edge ({\it edge} in {\it XSPEC}). 
Here we note that we have fixed the lower limit of the hydrogen column density 
($N_{\rm H}$) to $0.1\times10^{22}$ cm$^{-2}$, so that this parameter does not 
wander to an unphysically small value (NASA's HEASARC
nH tool gives $N_{\rm H} \approx 0.3\times10^{22}$ cm$^{-2}$ in the source direction).
We have also fixed the upper limit of the width ($\sigma_{\rm G}$; see Table 4)
of the Gaussian emission line to 1.0 keV during spectral fitting 
(as D' A\'i et al. 2006 did), so that this line does not
become unphysically broad. Moreover, as the lowest centroid energy of broad emission lines
found by Asai et al. (2000) was 5.9 keV, we have chosen this energy to be the lower limit
of our Gaussian line centroid energy.
From Table 3, we find that both the emission line and the absorption edge of the best fit model 
are significantly detected (see Fig. 4). With this knowledge,
we have, then, fitted the spectra of three additional data segments (one high intensity and two
low intensity). Table 4 shows the best fit parameter values from these spectral analyses,
as well as from the fitting of the spectrum of Table 3.
The emission line and the absorption edge significantly appear in
both the high intensity spectra. However, we do not find the blackbody component
and the emission line in the low intensity spectra, although the absorption edge component is
significantly present. The centroid energy of the emission line ($\sim 6$ keV) and
the threshold energy of the absorption edge ($\sim 8$ keV) indicate that these are
iron features (Asai et al. 2000; D' A\'i et al. 2006).

We note that, although the best fit parameters of the galactic ridge spectrum have
large errors (Table 2), the detection of the iron edge features is still certain.
This is because (1) the source was always $> 10-100$ times brighter than the galactic ridge,
and (2) the galactic ridge spectrum does not have an edge component.
The iron emission line from 1A 1744--361 was detected when the source was $\sim 100$ times
brighter than the galactic ridge. Therefore, it is unlikely that the detected iron emission line
originated from the ridge. Nevertheless, we have fitted the ridge spectrum with an 
alternative {\it XSPEC} model {\it wabs*(bremss+gauss)} in order to measure the strength
of the iron line. We find that photons cm$^{-2}$ s$^{-1}$ in this line is $(1.03\pm0.36)\times10^{-4}$
($\chi^2/{\rm dof} = 19.9/20$), while that in the iron emission line of the model 1
(high intensity data) of Table 4 is $(41.22^{+26.43}_{-16.67})\times10^{-4}$ ($\chi^2/{\rm dof} = 8.9/14$).
Therefore, from our results, and as the galactic ridge spectrum is separately
modeled, we conclude that the broad emission line and the absorption edge
originated from 1A 1744--361. 

\section {Discussion} \label{sec: 4}

In this paper, we have reported the correlated spectral and timing behavior of
the LMXB 1A 1744--361 for the first time, as well as the first discovery of iron 
features (broad emission line and absorption edge) in the energy spectra from this source.
We have estimated the luminosity range (in the unit of Eddington luminosity)
of  1A 1744--361, calculated the color-color diagram, hardness-intensity diagram and 
power spectra of this source, and found that 1A 1744--361
shows atoll behavior during outbursts. The source was observed at low luminosity
during the weak outburst in 2004, and at high luminosity
during the strong outbursts in 2003 and 2005. During the 2004 outburst, 1A 1744--361
was in the low-hard state, but there were not enough pointed
observations to determine whether the source was in such a state during
the rise or decay of 2003 and 2005 outbursts. 
Our analysis suggests that the dipper QPO (found by Bhattacharyya et al. 2006) is connected
to the low-hard state (AHB; see Figs. 2 and 3) of the source. This is 
consistent with the finding of such a QPO from the LMXB 4U 1746--37 only during the low-hard state
(Jonker et al. 2000). Also note that the dipper QPO
from the LMXB EXO 0748--676 was observed during the low intensity state, but not during
the high intensity state (Homan et al. 1999). The energy dependent dips of 1A 1744--361 
(discovered by Bhattacharyya et al. 2006) were observed only during the high intensity 
states of the source (see Fig. 2). This indicates that for 1A 1744--361 the dips are correlated with
the source states.

Spectral analyses show that the continuum part of the source spectra are well described
by a Comptonization plus blackbody (high intensity) or by a Comptonization (low intensity).
The Comptonization component might originate from an extended corona, while the origin
of the blackbody might be the neutron star surface and/or the inner accretion disk
(D' A\'i et al. 2006; White et al. 1986; Church \& Baluci\'nska-Church 2004). 
However, we note that some other model components (such as a bremsstrahlung, or a cutoff-powerlaw)
instead of Comptonization also give reasonable fits for most of the cases.
Nevertheless, we have used a simple Comptonization model 
for all our analyses, because a Comptonization component
was likely to be present in the spectrum (D' A\'i et al. 2006), and with the
first discovery of a broad emission line and an absorption edge in the 1A 1744--361 spectra,
we have primarily focused on these features.

The threshold energies of the absorption edge are within the range $7-9$ keV (Table 4), and hence are
consistent with those expected from ionized iron (D' A\'i et al. 2006). The emission line
is broad, and such broad iron lines (near 6 keV) are observed from many LMXBs, including dippers (Asai et
al. 2000). These lines may be broadened by either Doppler effects
due to Keplerian motion in the inner accretion disk, or Compton scattering in
disk corona (Asai et al. 2000). Analysis of high resolution spectra can likely
determine the source of this broadening by measuring the detailed
structure of these lines. If the Doppler effect is determined to be
the real cause, then the shapes and widths of these lines may be used
to constrain the inner edge radius of the disk, as well as the
Keplerian speed at that radius. The former can give an upper limit of
the neutron star radius (as the disk inner edge radius must be greater
than or equal to the stellar radius), while both quantities may be 
utilized to constrain the stellar mass. Note that the constraints on
neutron star mass and radius can be useful to understand the nature of
the high density cold matter at the stellar core, which is a
fundamental problem of physics (e.g., Lattimer \& Prakash 2001; 
Bhattacharyya et al. 2005). The centroid
energy of the broad iron emission line observed from 1A 1744--361 is less than 6.4 keV,
which could be due to unresolved iron absorption lines near $\sim 7$ keV (Parmar et al. 2002).
This suggests that 1A 1744--361 might show spectral absorption lines when observed
with higher spectral resolution missions (e.g., {\it Chandra}, {\it XMM-Newton} and {\it Suzaku}).

\acknowledgments

This work was supported in part by NASA Guest Investigator grants.

{}

\clearpage
\begin{deluxetable}{ccccccc}
\tablecolumns{7}
\tablewidth{0pc}
\tablecaption{Best fit parameters\tablenotemark{a} (with $1 \sigma$ error) for the low frequency 
(up to $\sim 100$ Hz) {\it RXTE} power spectra from 1A 1744--361.}
\tablehead{Reference & PLN $\nu$\tablenotemark{c} & PLN rms\tablenotemark{d} & L$_{f_{\rm
0}\tablenotemark{e}}$
& L$_{\lambda\tablenotemark{f}}$ & L$_{\rm rms\tablenotemark{g}}$ & $\chi^2/{\rm dof}$ \\
panel no.\tablenotemark{b} & & (\%) & (Hz) & (Hz) & (\%) & 
}
\startdata
1 & -- & -- & $2.43\pm0.13$ & $2.68\pm0.42$ & $13.1\pm0.7$ & $22.51/18$ \\
2 & $-1.50\pm0.21$ & $2.0\pm1.3$ & -- & -- & -- & $28.66/30$\\
3 & $-1.00\pm0.03$ & $6.7\pm0.2$ & -- & -- & -- & $23.26/30$\\
4 & $-0.94\pm0.02$ & $6.5\pm0.2$ & -- & -- & -- & $25.60/24$
\enddata
\tablenotetext{a}{Power spectra are fitted either by constant+Lorentzian, or by
constant+powerlaw in the energy range $\sim 2.6-30$ keV.}
\tablenotetext{b}{No. of the panel in Fig. 3, that shows the power spectrum and the position of the time
segment in the color-color diagram.}
\tablenotetext{c}{Index of power law ($\propto f^{\nu}$; $f$ is frequency) noise.}
\tablenotetext{d}{RMS of power law; lower limit of integration is 0.004 Hz.}
\tablenotetext{e}{Centroid frequency of Lorentzian ($\propto \lambda/[(f-f_{\rm
0})^2+(\lambda/2)^2]$).}
\tablenotetext{f}{Full width at half maximum (FWHM) of Lorentzian.}
\tablenotetext{g}{RMS of Lorentzian.}
\end{deluxetable}

\clearpage
\begin{deluxetable}{ccc}
\tablecolumns{3}
\tablewidth{0pc}
\tablecaption{Best fit model\tablenotemark{a} parameters for the $\sim 3.4-14$ keV galactic ridge spectrum ({\it
RXTE} PCA) from the position of 1A 1744--361 (when this source was not in outburst).}
\tablehead{Model Component & Parameter & value\tablenotemark{b}}
\startdata
Absorption & $N_{\rm H}$ ($10^{22}$ cm$^{-2}$) & $2.5_{-2.5}^{+3.5}$ \\
Raymond-Smith & $kT$ (keV) & $1.8_{-0.3}^{+0.7}$ \\
Power law & Photon index & $1.4_{-0.8}^{+0.9}$
\enddata
\tablenotetext{a}{{\it wabs(raymond+powerlaw)} model in {\it XSPEC}.}
\tablenotetext{b}{Best fit parameters with $1 \sigma$ errors. Reduced $\chi^2 =
15.8/21$. Flux $ = 2.2\times10^{-11}$ ergs cm$^{-2}$
s$^{-1}$ ($3.4-14$ keV).}
\end{deluxetable}

\clearpage
\begin{deluxetable}{clc}
\tablecolumns{3}
\tablewidth{0pc}
\tablecaption{Fitting of high intensity energy spectrum (ObsID 80431-01-02-02; {\it RXTE} PCA) from
1A 1744--361 with various spectral models of {\it XSPEC}.}
\tablehead{Model no. & Model\tablenotemark{a} & $\chi^2/{\rm dof}$}
\startdata
1 & {\it `ridge'+wabs*compTT} & $\frac{2084.8}{21} = 99.3$  \\
2 & {\it `ridge'+wabs*(compTT+bbodyrad)} & $\frac{764.9}{19} = 40.3$\tablenotemark{b}  \\
3 & {\it `ridge'+wabs*(compTT+bbodyrad+gauss)} & $\frac{27.37}{16} = 1.7$  \\
4 & {\it `ridge'+wabs*edge*(compTT+bbodyrad+gauss)} & $\frac{8.9}{14} = 0.6$\tablenotemark{c} 
\enddata
\tablenotetext{a}{The {\it `ridge'} model is the best fit model for galactic ridge spectrum (Table
2), and the parameters are frozen to the best fit values. For the {\it compTT}
model, spherical geometry is chosen.}
\tablenotetext{b}{See Fig. 4.}
\tablenotetext{c}{The probability (calculated from F-test) of the decrease of 
$\chi^2/{\rm dof}$ by chance from the value of the previous row to the that of the current row
is $3.7\times10^{-4}$.}
\end{deluxetable}

\clearpage
\begin{deluxetable}{ccccc}
\tablecolumns{5}
\tablewidth{0pc}
\tablecaption{Best fit parameters (with $1 \sigma$ error) for the {\it RXTE} PCA energy
spectra\tablenotemark{a} from 1A 1744--361.}
\tablehead{Model & 1 & 2 & 3 & 4\\
parameters &  &  &  & }
\startdata
$N_{\rm H}$\tablenotemark{b} & $0.10_{-0.00}^{+1.69}$ & $0.10_{-0.00}^{+0.42}$ &
$2.00_{-0.72}^{+0.69}$ & $0.51_{-0.41}^{+0.99}$ \\
$E_{\rm edge}$\tablenotemark{c} & $8.43_{-0.13}^{+0.30}$ & $8.38_{-0.19}^{+0.23}$ &
$8.68_{-0.32}^{+0.39}$ & $7.89_{-0.18}^{+0.20}$ \\
$D_{\rm edge}$\tablenotemark{d} & $0.12_{-0.05}^{+0.03}$ & $0.07_{-0.03}^{+0.02}$ &
$0.16_{-0.04}^{+0.04}$ & $0.18_{-0.05}^{+0.05}$  \\
$T_{\rm C}$\tablenotemark{e} & $3.50_{-0.81}^{+2.38}$ & $3.41_{-0.45}^{+1.03}$ &
$4.09_{-0.47}^{+0.87}$ & $3.55_{-0.17}^{+0.38}$ \\
$\tau_{\rm C}$\tablenotemark{f} & $9.53_{-2.28}^{+1.94}$ & $9.11_{-1.55}^{+1.05}$ &
$11.58_{-1.61}^{+1.40}$ & $14.01_{-1.59}^{+0.91}$ \\
$T_{\rm BB}$\tablenotemark{g} & $1.31_{-0.06}^{+0.03}$ & $1.26_{-0.07}^{+0.05}$ &
-- & --  \\
$E_{\rm G}$\tablenotemark{h} & $5.97_{-0.07}^{+0.09}$ & $5.90_{-0.00}^{+0.11}$ & 
-- & --  \\
$\sigma_{\rm G}$\tablenotemark{i} & $0.62_{-0.26}^{+0.27}$ & $0.70_{-0.29}^{+0.11}$ & 
-- & --  \\
\hline
$EW$\tablenotemark{j} & $117_{-47}^{+75}$ & $131_{-64}^{+51}$ & -- & --  \\
Flux\tablenotemark{k} & $2.2\times10^{-9}$ & $2.1\times10^{-9}$ & $3.1\times10^{-10}$ & $2.9\times10^{-10}$ \\
$\frac{\chi^2}{{\rm dof}}$ & $\frac{8.9}{14}$ & $\frac{13.9}{10}$ & $\frac{13.1}{19}$ &
$\frac{11.7}{19}$
\enddata
\tablenotetext{a}{1: ObsID 80431-01-02-02 ({\it plus} symbols of panel {\it d} of Fig. 3);
2: ObsID 80431-01-02-04 ({\it plus} symbols of panel {\it c} of Fig. 3);
3: ObsID 90058-04-01-00 ({\it diamond} symbols of Fig. 2);
4: ObsID 90058-04-02-00 ({\it cross} symbols of Fig. 2). First two spectra (high intensity) are fitted with
{\it `ridge'+wabs*edge*(compTT+bbodyrad+gauss)} model of {\it XSPEC} ({\it `ridge'} model is described
in Table 3), while the last two spectra (low intensity) are fitted with {\it
`ridge'+wabs*edge*compTT} model.}
\tablenotetext{b}{Hydrogen column density ($10^{22}$ cm$^{-2}$) from {\it wabs}
model in {\it XSPEC}; imposed lower limit is $0.1$.}
\tablenotetext{c}{Threshold energy (keV) of the edge ({\it edge} model in {\it XSPEC}).}
\tablenotetext{d}{Absorption depth at the threshold of the edge.}
\tablenotetext{e}{Temperature (keV) of the Comptonizing plasma ({\it compTT} model in {\it XSPEC}).}
\tablenotetext{f}{Optical depth of the Comptonizing plasma.}
\tablenotetext{g}{Blackbody temperature (keV) ({\it bbodyrad} model in {\it XSPEC}).}
\tablenotetext{h}{Centroid energy (keV) of Gaussian emission line ({\it gauss} model in {\it
XSPEC}); imposed lower limit is $5.9$.}
\tablenotetext{i}{Width (keV) of Gaussian emission line; imposed upper limit is $1.0$.}
\tablenotetext{j}{The equivalent width (eV) of Gaussian emission line.}
\tablenotetext{k}{Flux in ergs cm$^{-2}$ s$^{-1}$ ($3.4-14$ keV).}
\end{deluxetable}

\clearpage
\begin{figure}
\epsscale{0.8}
\plotone{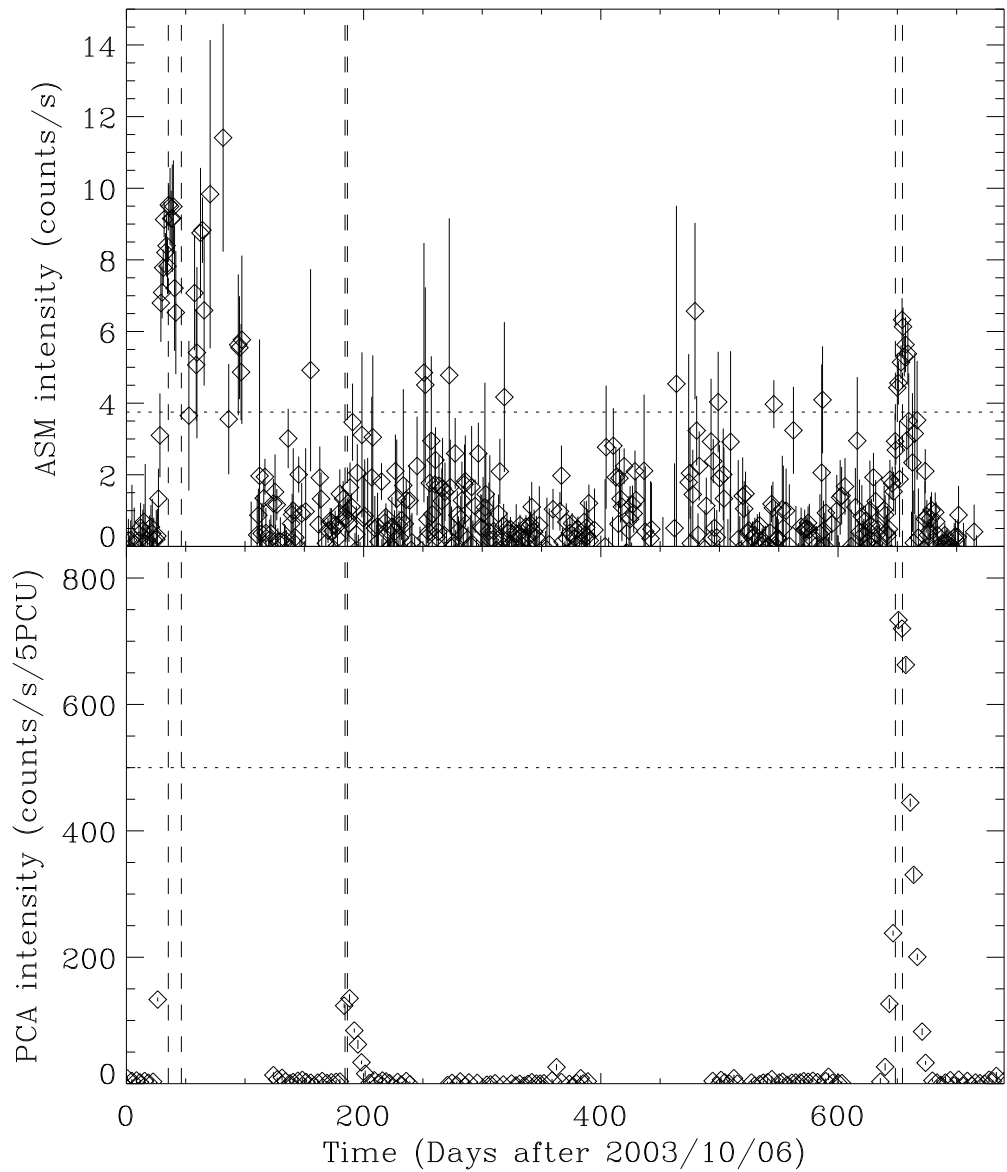}
\caption {The long term ASM and PCA (galactic bulge scan) light curves of 1A 1744--361. The pairs of dashed
vertical lines show the time intervals of pointed {\it RXTE} observations of the source during
the outbursts in 2003, 2004 and 2005. Each of these intervals contains several observations.
The dotted horizontal lines give the 50 mCrab intensity
level. Note that the PCA galactic bulge scan was prevented due to the angular proximity to the sun 
during the first outburst, which caused a data gap. We also note that 
1A 1744--361 was probably not detected with the PCA in the quiescent state, and outburst activity
is rare, as the PCA galactic bulge scan did not find any outburst from this source
during the years $1998-2002$.
}
\end{figure}

\clearpage
\begin{figure}
\epsscale{1.0}
\plotone{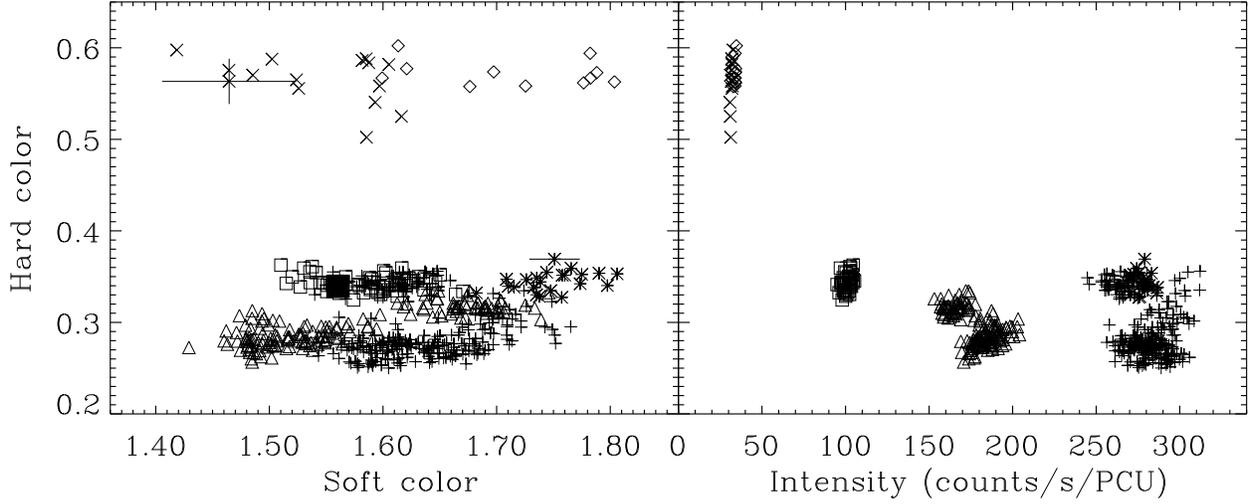}
\caption {Color-color diagram (left panel) and hardness-intensity diagram (right panel)
of 1A 1744--361 using the {\it RXTE} PCA pointed observation data of the years 
2003, 2004 and 2005. The definitions of the colors and the energy range of
the intensity are given in \S~2. Here we use the data only from the top Xenon layers. 
Each point is for 64 s of data. 
The {\it cross} symbols are for the ObsID 90058-04-02-00 (2004 data that show a $\sim 2.5$
Hz QPO, and a possible kHz QPO (Bhattacharyya et al. 2006)), the {\it diamond} symbols are
for ObsID 90058-04-01-00 (2004 data that show a $\sim 3.5$ Hz QPO (Bhattacharyya et al.
2006)), the {\it square} symbols are for the ObsID 91050-05-01-00 (2005 data that show a
thermonuclear X--ray burst (Bhattacharyya et al. 2006)), the {\it triangle} symbols are for
the rest of the 2005 data, the {\it star} symbols are for the two segments of the ObsID
80431-01-02-00 (2003 data that show energy dependent dips (Bhattacharyya et al. 2006);
but the dip portions are excluded), and the {\it plus} symbols are for the rest of the
2003 data. The {\it filled square} symbol is for the data just before the burst.
Here we exclude the time intervals of obvious flares, burst and dips.
Two sets (one for low intensity data and another for high intensity data) of typical 
$1 \sigma$ error bars are shown in the color-color diagram.
The ranges of soft color, hard color and intensity strongly indicate 
the atoll nature of 1A 1744--361 during the outbursts (see \S~2).
}
\end{figure}

\clearpage
\begin{figure}
\epsscale{0.9}
\plotone{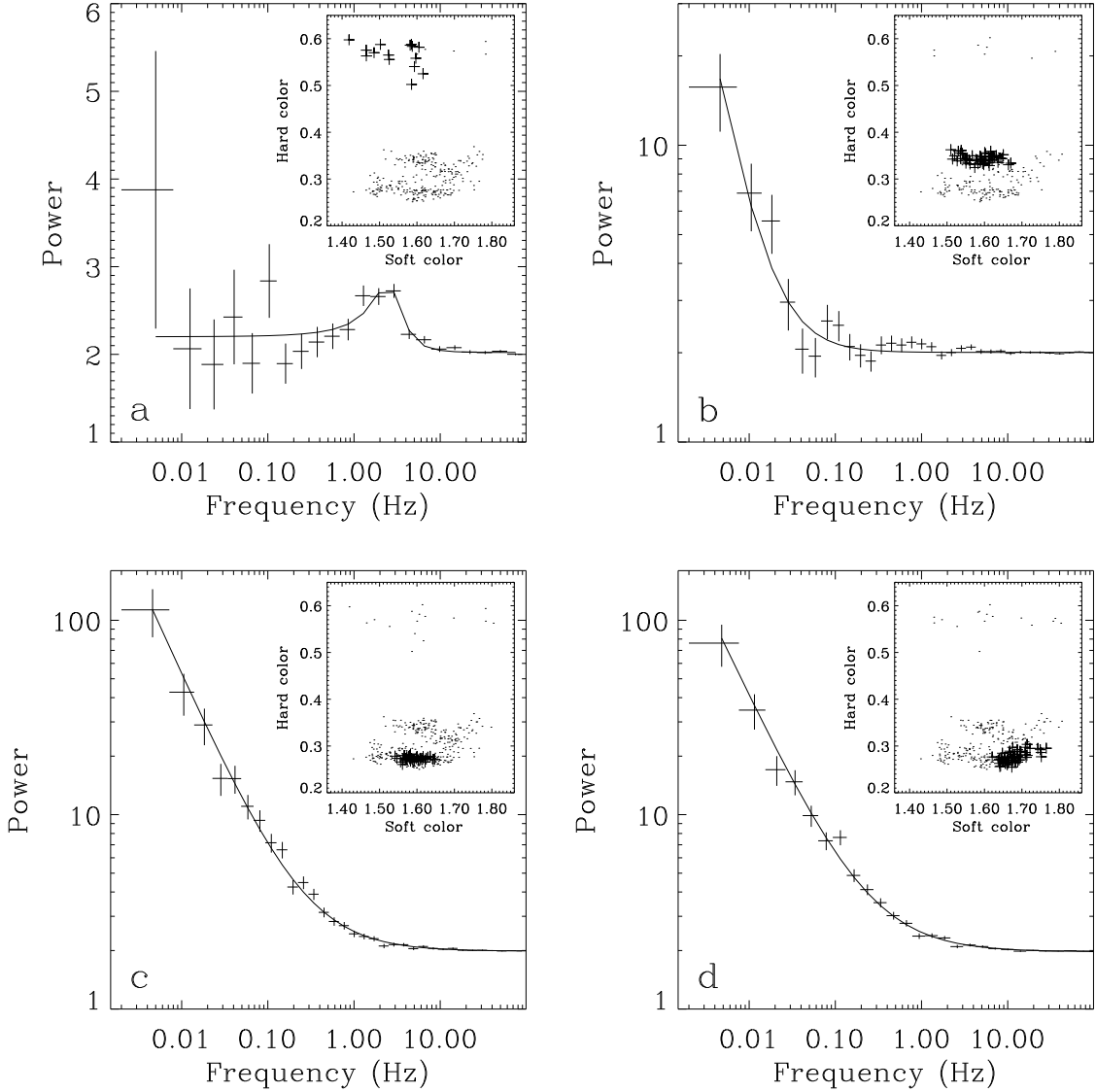}
\caption {Low frequency power spectra (for the energy range $\sim 2.6-30$ keV) of
1A 1744--361
using {\it RXTE} PCA data. Panels {\it a}, {\it b}, {\it c} and {\it d} are for the ObsIDs
90058-04-02-00, 91050-05-01-00, 80431-01-02-04 and 80431-01-02-02 respectively.
For each panel, the main panel shows the data points and the best fit model (solid line;
see Table 1). The horizontal lines around the data points give the frequency bin, and the
corresponding vertical lines give the $1 \sigma$ errors of powers. Each inset panel
shows the color-color diagram (same as in Fig. 2) and the {\it plus} symbols show the data used to
calculate the power spectrum in the corresponding main panel. Panel {\it a} is for the low
intensity AHB (see \S~2) state of the source, and the power spectrum is well fitted with a
constant+Lorentzian model. The `constant' describes the Poisson noise level and the 
Lorentzian describes the dipper QPO (Bhattacharyya et al.  2006). Panels {\it b}, {\it c} and 
{\it d} are for the banana state of 1A 1744--361, and a constant+powerlaw model fits the
power spectra well. Here the `powerlaw' describes the very low frequency noise (VLFN).
}
\end{figure}

\clearpage
\begin{figure}
\vspace{-0.5cm}
\epsscale{0.8}
\plotone{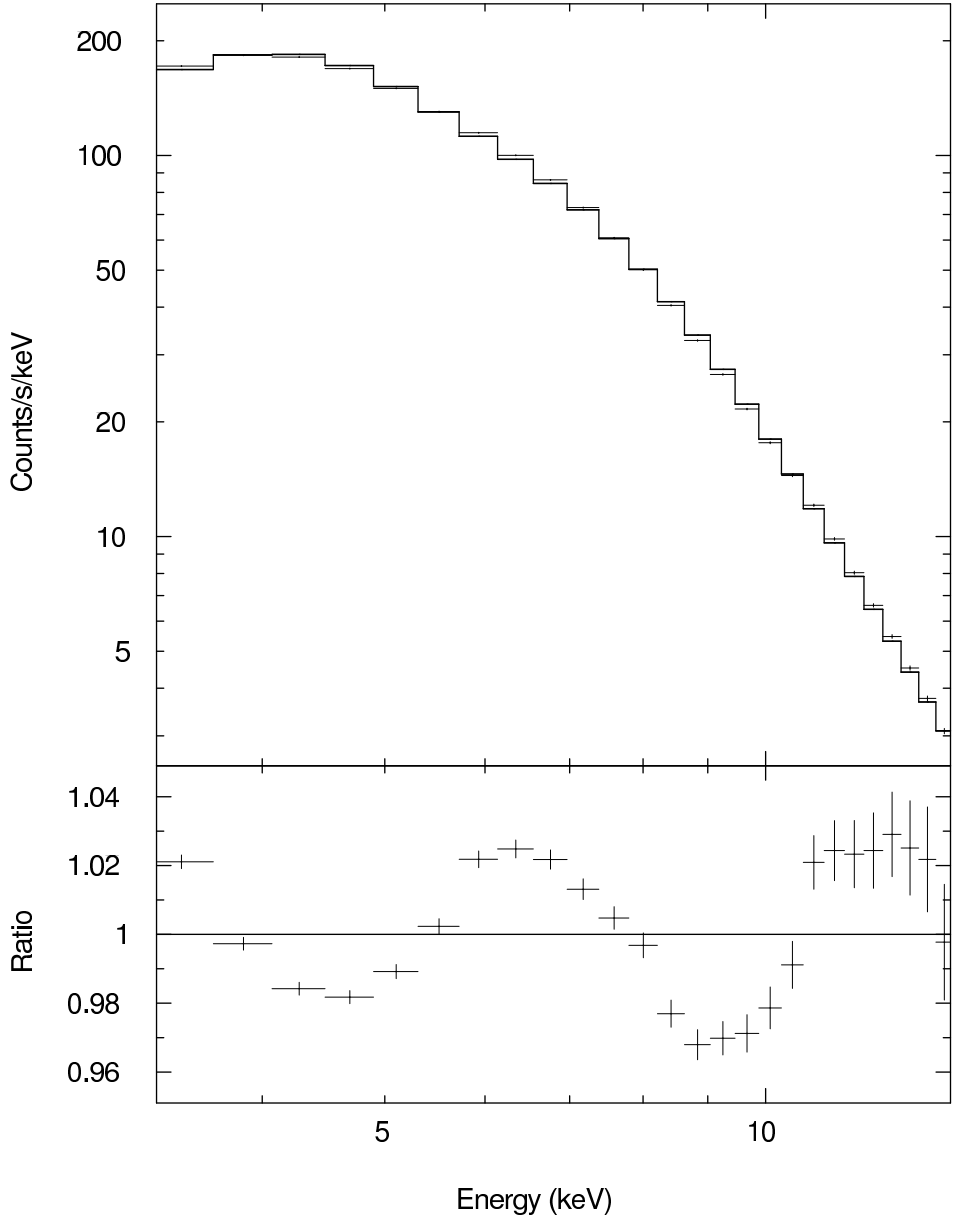}
\vspace{-0.2cm}
\caption {{\it RXTE} PCA energy spectrum from 1A 1744--361 for the ObsID 80431-01-02-02
(time segment denoted by {\it plus} symbols of the inset panel of panel {\it d} of Fig. 3; high intensity banana state).
Here we use only the top Xenon layers, and fit the data with
a continuum model ({\it `ridge'+wabs*(compTT+bbodyrad)} in {\it XSPEC}; model no. 2 of Table 3). 
The {\it `ridge'} model is the best fit model ({\it wabs*(raymond+powerlaw)} in {\it XSPEC}) for galactic
ridge emission and we fix the parameters of this model to the best fit value (Table 2).
The upper panel shows the data
points and the model (solid histogram). The lower panel shows the data to model ratio.
The lower panel suggests that an additional broad emission line model component (near $\sim 6$ keV)
and an additional absorption edge model component (near $\sim 8$ keV) are
required for a good fit.
}
\end{figure}

\end{document}